\documentclass[12pt]{article}
\usepackage{amssymb}

\begin{document}

\title{Comment on ``Removable singularities for solutions of coupled Yang-Mills-Dirac equations" [\emph{J. Math. Phys.} \textbf{47}, 103502 (2006)]}
\author{Thomas H. Otway\thanks{%
email: otway@yu.edu} \\
\\
\textit{Department of Mathematics, Yeshiva University,}\\
\textit{\ \ New York, New York 10033}}
\date{}
\maketitle

\begin{abstract}
A lemma from elliptic theory is used to improve a recent result by
Li concerning the removability of an isolated point singularity
from solutions of the coupled Yang-Mills-Dirac equations. MSC2000:
35J60, 70S15
\end{abstract}

\bigskip

Recently, Wei Li proved \cite{L} that a smooth solution of the
coupled Yang-Mills-Dirac equations in a punctured Euclidean 4-disk
$B\backslash\{0\}$ is equivalent via a continuous gauge
transformation to a smooth solution in all of $B$ whenever $F\in
L^2(B)$ and $\phi\in L^{8/3}(B).$ Here $F$ denotes the Yang-Mills
field and $\phi$ is the spinor field. This represents a technical
improvement of a well known theorem by Parker \cite{P}, in that no
condition is placed on the derivative of the spinor field in
\cite{L}. An important feature of Li's proof is that the fields
are estimated by purely analytic arguments, away from the
singularity, in terms of conformally invariant norms; an
application of Uhlenbeck's broken Hodge gauges \cite{U} completes
the proof.

In fact the $L^p$ hypothesis on $\phi$ in \cite{L} can be weakened
by an argument which is also purely analytic. A lemma due to L. M.
Sibner \cite{Si} can be used to show the sufficiency of the
assumption $\phi\in L^p(B),$ $p>2,$ provided the other hypotheses
of \cite{L} are retained.

\bigskip

\textbf{Lemma 1} (L. M. Sibner). \emph{Let the nonnegative scalar
function $u$ be $C^\infty$ in the punctured $n$-disk
$B\backslash\{0\},$ for $n>2,$ and satisfy there the subelliptic
inequality}
\begin{equation}\label{scal}
    \Delta u + g(x) u \geq 0
\end{equation}
\emph{for a function $g\in L^{n/2}(B).$ If for $1/2<q_0<q$ we have
$u\in L^{2nq_0/(n-2)}(B)\cap L^{2q}(B),$ then
$\nabla\left(u^q\right)\in L^2(B)$ and in a sufficiently small
$n$-disk $\tilde B,$}
\[
    \int_{\tilde B}\eta^2|\nabla\left(u^q\right)|^2\,d(vol)\leq
C\int_{\tilde B}|\nabla \eta|^2u^{2q}\,d(vol)
\]
\emph{for a positive constant $C$ and all $\eta \in
C^\infty_0(\tilde B).$}

\bigskip

The proof of Lemma 1 in \cite{Si} depends on the use of a delicate
test function introduced by Serrin (\cite{Se}, see also \cite{GS},
Sec.\ 3), but is otherwise elementary.

Thus we have:

\bigskip

\textbf{Theorem 2}. \emph{Let the pair $\left(F,\phi\right)$
smoothly satisfy the coupled Yang-Mills-Dirac equations in the
punctured Euclidean 4-disk $B\backslash\{0\}.$ If $F\in L^2(B)$
and $\phi\in L^p(B)$ for some $p>2,$ then $F$ and $\phi$ are
equivalent via a continuous gauge transformation to a smooth
solution over all of $B.$}

\bigskip

\emph{Proof}. In Lemma 1, choose $u=|\phi|$ and $g=k|F|$ for
constant $k.$ Then (\ref{scal}) is satisfied, by Proposition 3.3
of \cite{L}. Choose $q_0=1/\left(2-\varepsilon\right),$ for
$0<\varepsilon<2,$ and
$q=n/\left(2-\varepsilon\right)\left(n-2\right).$ Lemma 1 implies
that $\nabla\left( |\phi|^{\,q}\right)\in L^2(B).$ Applying the
Sobolev Theorem, we conclude that $|\phi|\in L^r(B)$ for
$r=2\left[n/\left(n-2\right)\right]^2/\left(2-\varepsilon\right).$
If $n=4,$ then $r$ exceeds 8/3. Now apply Theorem 4.1 of \cite{L}.
This completes the proof of Theorem 2.

\bigskip

The $L^p$ estimate of $\phi$ in the proof of Theorem 2 does not
use any of the properties of a geometric object. In particular, it
does not use the conformal weight of $\phi.$ Iterating this
estimate will show that $\phi$ lies in the space $L^r$ for any
finite value of $r$ (see \cite{O}, Proposition 3.7); but the
arguments of \cite{L} make such an iteration unnecessary.

I am grateful to Yisong Yang for drawing my attention to \cite{L}.

\end{document}